\documentstyle[aps,prl,epsf]{revtex} 
\begin{document}
\title{Edge critical behaviour of the 2-dimensional
tri-critical Ising model}
\author{Ian Affleck}
\address{Institute for Theoretical Physics, University of
California, Santa Barbara, CA93106-4030, U.S.A. and 
Canadian Institute for Advanced
Research  and Department of Physics and Astronomy, University of 
British 
Columbia, 
Vancouver, B.C., Canada, V6T 1Z1}
\maketitle
\begin{abstract}
Using previous results from boundary conformal field theory
and integrability, a phase diagram is derived for the 2 dimensional 
Ising model at its bulk tri-critical point as a function of boundary
magnetic field and boundary spin-coupling constant.  A boundary 
tri-critical point separates phases where the spins on the boundary
are ordered or disordered.  In the latter
range of coupling constant, there is a non-zero critical field
where the magnetization is singular.  In the former range, as the
temperature is lowered, the boundary undergoes a first order
transition while the bulk simultaneously undergoes a second order transition.
\end{abstract}
Conformal field theory has led to many exact results on 2-dimensional 
dimensional  critical pheomena both with regard to bulk behaviour
and edge or boundary behaviour.  (For a review see \onlinecite{DiF}.)
Assuming the bulk system is at a critical point, one can consider
critical behaviour at the boundary as a function of various fields
and interactions applied near the boundary.  In general, various
boundary phases and critical points exist for a given bulk
critical point.  These models can be used to describe either
2-dimensional classical systems at bulk critical points or
else semi-infinite quantum chains at zero temperature.  Some of
these latter systems find experimental application to strongly
correlated electron impurity problems.    While the boundary
phase diagram of the critical Ising model is well understood,\cite{Cardy} 
surprisingly, this is not so for the next simplest case,
the tri-critical Ising model. 
Six conformally invariant boundary conditions
have been constructed using the fusion method by Cardy\cite{Cardy} which
should correspond to boundary critical points.  Certain integrable
renormalization group (RG) flows between these critical points have
been constructed by Chim \cite{Chim}.  The purpose of this note is simply
to connect the points with a phase diagram written in terms of
microscopic parameters.  This is shown schematically in
Fig. 1.  The most surprising conclusion is perhaps the
existence of a phase, in zero magnetic field,
 where the spins on the boundary exhibit long range
order while those in the bulk do not. This is the 
physical interpretation of a boundary
condition for which the corresponding boundary state is
a sum  of boundary states corresponding to a spin up
or spin down boundary condition.  
\begin{figure}
\epsfxsize=8 cm
\centerline{\epsffile{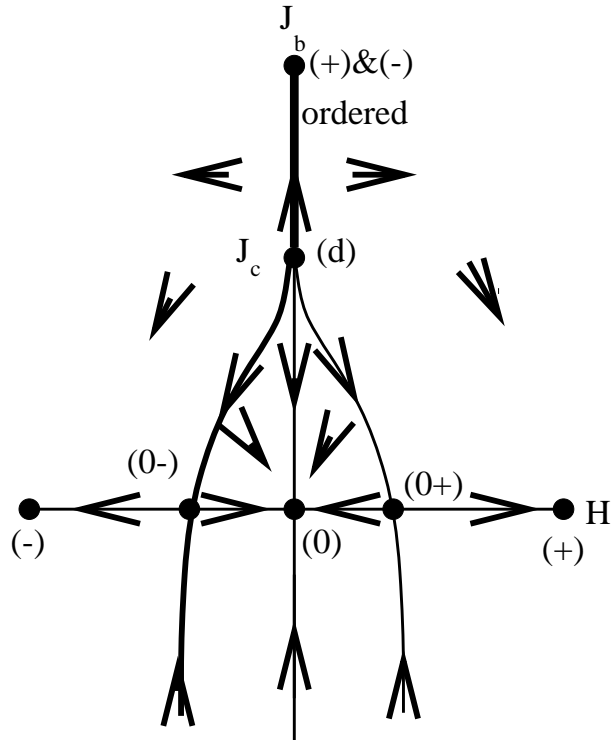}}
\caption{Schematic phase diagram of the boundary tri-critical
Ising model. Arrows 
indicate direction
of RG flows as the length scale is increased.  Along the thick
line the spins on the boundary are ordered.}
\label{fig:pd}
\end{figure}

We first briefly review the simpler case of the ordinary 
critical Ising model.  In that case, Cardy identified only
3 conformally invariant boundary conditions corresponding
to spin up, spin down and free.  There are no relevant
boundary operators at the spin up/down critical points,
indicating that they are stable against the addition to the Hamiltonian
of arbitrary perturbations located near the boundary.   
On the other hand, the free boundary condition has one
relevant operator, of dimension $x=1/2$.  (Boundary operators are
relevant if they have scaling dimension $x<1$.)  The corresponding
relevant coupling constant is naturally interpreted as a boundary
magnetic field.  By standard scaling arguments we expect the
boundary magnetization, to scale with boundary field as:
\begin{equation} |m| \propto |H|^{x/(1-x)}\propto |H|.\label{ordis}
\end{equation}
This behaviour is expected to be independent of the details of
the spin coupling constants near the boundary.  (Only the
constant of proportionality in Eq. (\ref{ordis}) will depend on
these coupling constants, not the exponent.) No additional
critical points are expected at any finite field.  
The renormalization
group flow between free and fixed b.c.'s (boundary conditions)
was shown to be integrable by Ghoshal and Zamalodchikov in a
pioneering paper \cite{Ghoshal}.

The phase diagram for the tri-critical Ising
model is more interesting.  This model can be defined as a
spin-1 Ising model with a crystal field term which favours
the  $S=0$ state over $S=\pm 1$ or equivalently a diluted
spin-1/2 Ising model.  The (classical) Hamiltonian is:
\begin{equation}
 H=-J\sum_{<i,j>}S_iS_j +\mu\sum_iS_i^2,\ \  (S_i=-1,0,1).
 \label{bulkHam}\end{equation}
 The schematic (bulk) phase diagram is drawn in Fig. 2.  There
 is a second order phase transition line in the Ising
 universality class and also a first order phase
 transition line separating phases with unbroken and broken symmetry.  These
 lines join at a tri-critical point.  By an approximate
 transfer matrix mapping, one can show that this classical
 model exhibits the same critical behaviour as a quantum
 chain at T=0.  This model has the Hamiltonian:
 \begin{equation}
 H=-\sum_i[S^z_iS^z_{i+1}-D(S^z_i)^2+H_TS^x_i].\end{equation}
 $S^a_j$ now label quantum S=1 operators.  The transverse field
  $H_T$ now controls the  temperature in
  the classical model.
  While the ordinary Ising critical point corresponds to the
  simplest conformal field theory with central charge, c=1/2,
  the tri-critical point corresponds to the next unitary
  minimal model with c=7/10.
  
  \begin{figure}
\epsfxsize=10 cm
\centerline{\epsffile{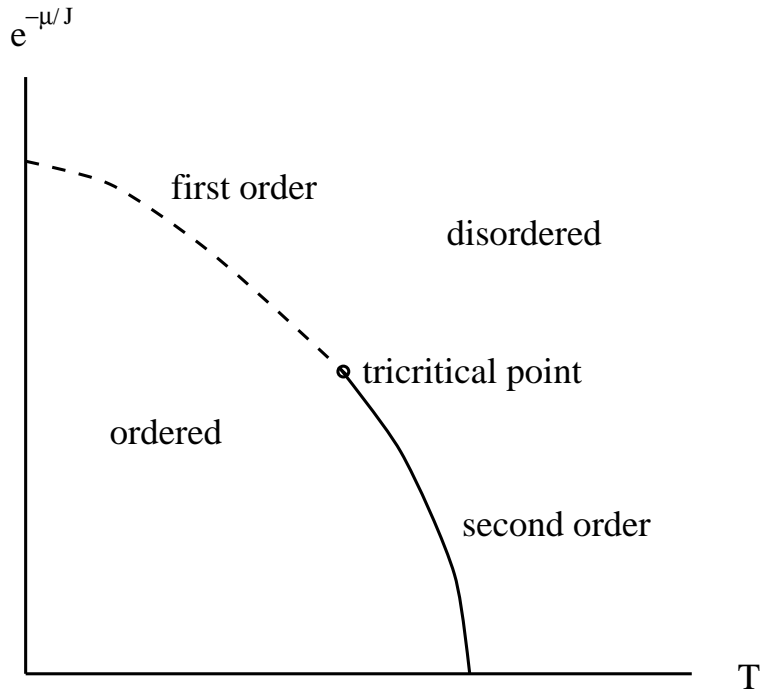}}
\caption{Schematic bulk phase diagram of the spin-1 Ising model of Eq. 
(\protect\ref{bulkHam}).}
\label{fig:bulk}
\end{figure}
  
  We now wish to consider the classical model on a semi-infinite half-plane
  with a boundary as shown in Fig. 3. 
   Although various microscopic boundary
  interactions could be considered, for our purposes it is
  enough to consider a boundary field, $H$ and a modified interaction,
  $J_b$ along the boundary. These couplings are indicated in
  Figure 3.  In the corresponding quantum chain, the Hamiltonian is:
  \begin{equation}
  H=-\sum_{i=0}^\infty S^z_iS^z_{i+1}+
  \sum_{i=1}^\infty [-H_TS^x_i+D(S^z_i)^2]-H_{Tb}S^x_0
  +D_b(S^z_0)^2-HS^z_0.
  \end{equation}
  Roughly speaking, increasing the boundary interaction, $J_b$, in
  the classical model corresponds to decreasing $|H_{Tb}|$ 
   and $D_b$ in the 
  quantum model thus enhancing the tendency for the spins
  to order at the boundary.
  
  \begin{figure}
\epsfxsize=10 cm
\centerline{\epsffile{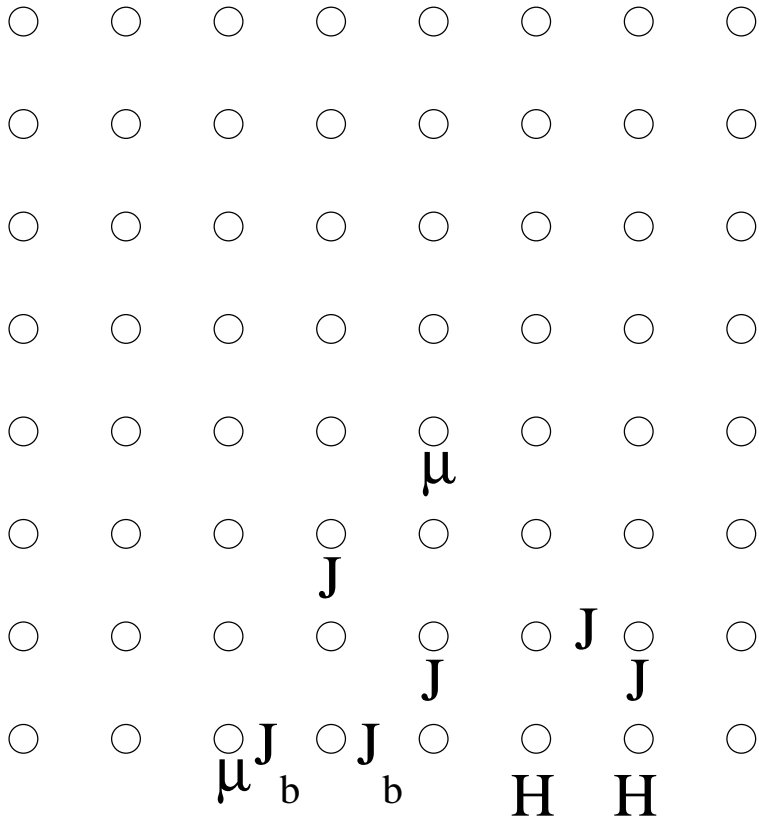}}
\caption{Couplings and field for the boundary tri-critical Ising model.
The bulk parameters, $J$ and $\mu$ are adjusted to the tri-critical point.
The magnetic field is applied only at the boundary.}
\label{fig:couplings}
\end{figure}
  
  The phase diagram in Figure 1 can be deduced rather straightforwardly
  from the properties of the 6 conformally invariant boundary
  states found by Cardy and discussed by Chim, and  from the 
  integrable RG flows discussed by Chim.  We first review these
  boundary states.  Using the fusion approach, one finds
  6 boundary states corresponding to the 6 primary fields in
  the (bulk) tri-critical Ising model.  Their physical properties
  have been elucidated, to some extent, by Chim, and we use
  his notation for them.  There are 2 states corresponding
  to spin up and spin down boundary conditions, $(\pm )$.  There are
  2 more boundary conditions, $(0\pm )$ which also break the $Z_2$ symmetry
  but appear to have the spins at the boundary only partially
  polarized.  [Chim actually labelled the negative polarization
  b.c. $(-0)$ rather than $(0-)$ but we prefer the latter
  notation.]  $(0)$ is the free boundary condition \cite{Saleur}.
    Finally there is
  one more b.c. which does not break the $Z_2$ symmetry and is
  labelled $(d)$ (for degenerate).  The correspondance between
  the fusion label and the physical label for the corresponding
  boundary states is:
  \begin{eqnarray}
  |\tilde 0>&=&|(-)>\nonumber \\
  |\tilde {3\over 2}>&=&|(+)>\nonumber \\
  |\tilde {1\over 10}>&=&|(0-)>\nonumber \\
  |\tilde {3\over 5}>&=&|(0+)>\nonumber \\
  |\tilde {7\over 16}>&=&|(0)>\nonumber \\
  |\tilde {3\over 80}>&=&|(d)>.
  \end{eqnarray}
  By the standard fusion rules, the (primary) boundary operator content
  with the boundary condition corresponding to the state
  $|\tilde a>$ is the set of operators appearing in the (bulk)
  operator product expansion (OPE) of ${\cal O}_a\times {\cal O}_a$.
  It thus follows that the $(\pm )$ b.c.'s admit no relevant
  operators; they are completely stable. (The only operator appearing
  in the OPE is the identity operator which just corresponds to
  the possibility of adding a c-number to the quantum Hamiltonian,
  having no effect on the critical behaviour.)
   The only primary
  operator at the free b.c. $(0)$ has dimension 3/2.  This follows
  from the OPE
  \begin{equation}\sigma '\times \sigma '=I + \epsilon '',\end{equation}
  where $\sigma '$ and $\epsilon ''$ are the primary fields of
  dimension 7/16 and 3/2 respectively.  Thus the free b.c. is
  also a stable fixed point!  This is a somewhat surprising result
  since it implies that adding a boundary magnetic field does not
  destabilize the free fixed point in the tri-critical Ising model, unlike
  what happens in the ordinary Ising model.  The partially polarized
  boundary conditions $(0\pm )$ have one relevant boundary operator
  with $x=3/5$.  Thus, there should be one unstable direction and
  one stable direction in the RG flow in the $(J_b,H)$ plane at the
  corresponding fixed points.  The
  $(d)$ b.c. has two relevant boundary operators of dimension
  $x=1/10$ and $3/5$, so both directions should be unstable
  at this fixed point.  Finally, it is important to consider the
  boundary condition labelled $(+) \& (-)$ by Chim.  The corresponding
  boundary state is
  \begin{equation}
  |(+) \& (-)>= |(+)>+|(-)>=|\tilde 0>+|\tilde {3\over 2}>.\end{equation}
  The corresponding OPE is:
  \begin{equation}
  [I+\epsilon '']\times [I+\epsilon '']=2[I+\epsilon ''].\end{equation}
  Thus there should be only one relevant boundary operator, of dimension
  $x=0$, at this critical point (disgarding the identity operator
  which is always present and has no effect). The presence of a
  non-trivial boundary operator with dimension 0 is the hallmark
  of an ordered phase, or equivalently 
  a first order phase transition with magnetic field.  It is natural to associate
  this operator with a boundary magnetic field.  The usual scaling
  law,
  \begin{equation}
  |m|\propto |H|^{x/(1-x)},\end{equation}
  implies $|m|\propto |H|^0$, i.e. a discontinuous jump in $m$ as
  $H$ passes through 0.  This in turn implies long range order
  in zero field.  It is also noteworthy that there are no
  additional relevant operators with the $  (+) \& (-)$ b.c.  Thus
  we expect it is stable against small variations of $J_b$ at $H=0$.
  This is different than the other combination
  $(0+)\& (0-)$ which has, in addition to an $x=0$ boundary operator,
  2 other relevant boundary operators with $x=2/5$.  It is
  also different than the situation in the ordinary Ising
  model where the combination of spin up and down gives 3 relevant
  boundary operators.   The identification of a boundary state
  which is a sum of two or more other boundary states with long range
  order was also made in the context of a critical line separating
  two semi-infinite Ising planes \cite{Oshikawa} and in the boundary
  3-state Potts model \cite{Affleck}.  However, in both those cases 
  it is an unstable fixed point, even in zero field.  The somewhat
  unusual feature of the tri-critical Ising model is that the broken
  symmetry phase is stable.  Related phenomena also occur in
  quantum Brownian motion on a triangular lattice \cite{tri}.
  
  It is now a relatively straightforward matter to connect the points
  to obtain the schematic phas diagram of Fig. 1. 
   Several comments
  about this phase diagram are in order. The flows from $(d)$ to
  $(0)$ and $(+)\&(-)$ and from $(0+)$ to $(0)$ and $(+)$
  are integrable. Since there is only one relevant operator at
  the $(0\pm )$ fixed points there must be lines in the $(J_b,H)$
  plane which flow towards them.  These lines must end at the
  tri-critical point $(d)$ since it is the only fixed point
  with two relevant operators.  
  The values
  of $J_b$ at the fixed points are in general unknown 
  (and are presumably {\it not} equal at 5 of the fixed 
  points as drawn in Fig. 1.).  However, we do expect that
  $J_b$ is smaller at $(0)$ than at $(d)$. It is natural to
  place the $(+)\&(-)$ fixed point at $J_b=\infty$ since there
  the spins along the boundary are perfectly ordered.  The values of
  $J_b$ at the $(\pm )$ and $(0\pm )$ fixed points are relatively
  insignificant and simply correspond to points where the leading
  irrelevant coupling constant vanishes.  Although a boundary
  field is irrelevant in the central phase of the phase diagram 
  this {\it does not} imply that the boundary magnetization is
  zero in the presence of a non-zero field.  Irrelevant operators
  will still lead to a non-zero magnetization which should be
  an {\it analytic} function of $H$, thus being linear at small
  $H$.  
  At the $(0\pm )$ fixed points, we expect $m(H)$ to be
  singular, behaving as:
  \begin{equation}
  m-m_c- a(H-H_c)\to b_{\pm}|H-H_c|^{x/(1-x)}=b_{\pm}|H-H_c|^{3/2},
  \end{equation}
  since $x=3/5$. Here the amplitudes, $b_{\pm}$ are presumably
  different for $H>H_c$ and $H<H_c$.  The linear term, $\propto a$,
  is non-singular.  
   This is a relatively mild singularity since
  both $m$ and its first derivative remain finite and continuous,
  while the second derivative diverges. 
  The shape
  of the phase boundary near $(d)$ is determined from the scaling
  dimension of  $H$ and $J_b-J_c$ to be:
  \begin{equation}
  J_c-J_b(H)\propto |H|^{4/9}.\end{equation}
  At the tri-critical point $(d)$, 
  \begin{equation}
  |m|\propto |H|^{1/9},\end{equation}
  and for $J_b>J_c$, $m$ has a first order jump at $H=0$.
  
  These results are all consistent with the ``g-theorem''\cite{Affleck1}
   that states
  that the groundstate degeneracy, $g$, always decreases during
  a boundary RG flow.  The $g$ values are given by:
  \begin{eqnarray}
  g_{(\pm)}&=&C \nonumber \\
  g_{(0\pm )}&=&C\eta^2 \nonumber \\
  g_{(0)}&=&\sqrt{2}C \nonumber \\
  g_{(d)}&=&\sqrt{2}\eta^2C\nonumber \\
  g_{(+)\&(-)}&=&2C,\end{eqnarray}
  where:
  \begin{equation}
  C=\sqrt{\sin {\pi \over 5}\over \sqrt{5}},\ \  \eta = 
  \sqrt{\sin {2\pi \over 5}\over \sin {\pi \over 5}}.
  \end{equation}
  Noting that $\eta^2\approx 1.61803>\sqrt{2}$
  we see that all
  RG flows in Fig. 1 are consistent with the $g$ theorem.  This was
  already observed by Chim in the special cases of the integrable flows.  
  
  The existence of the ordered line may seem somewhat surprising since
   there is long range order at finite temperature along the 
   (one dimensional) boundary,
  for $J_b>J_c$, even though the (two dimensional) bulk is disordered (or,
  more accurately, is sitting at a critical point separating
  ordered and disordered phases).  This is
  surely reasonable at $J_b\to \infty$ 
  but may be harder to swallow for finite $J_b$. 
  In the quantum chain context this behaviour is not so unfamiliar.  
  We may think of the $(+)\& (-)$ fixed point as corresponding
  to $H_{Tb}=0$.  In this limit $S^z_0$ commutes with the 
  Hamiltonian
  and there are 2 degenerate groundstates, with $S^z_0=\pm 1$. 
  (We assume that $D_b$ is sufficiently large and negative that
  these states have lower energy than the one with $S^z_0=0$.)
  These groundstates have non-zero, equal
  and opposite values of the magnetization, localized near the boundary.
    Applying an
  infinitesimal boundary field picks out one of these two groundstates,
  leading to the discontinuity.
  The above RG analysis
  implies that a small transverse boundary field is {\it irrelevant}
  so the jump in the magnetization at $H=0$ should persist for a range
  of non-zero $H_{Tb}$.  A somewhat related phenomena occurs in the
  {\it ferromagnetic} Kondo problem.  The Kondo coupling constant
  is irrelevant in this case so that the impurity spin decouples
  from the conduction electrons at the stable fixed point.  Since
  the impurity spin operator then commutes with the Hamiltonian, 
  there are degenerate groundstates and a discontinuous magnetization
  in an applied field.  
  
  It is interesting to consider varying $T$ through the bulk phase transition
  (with the Hamiltonian held fixed) .  Referring to Fig. 2, we see that for
  large $\mu$ the bulk transition is continuous and  in the usual Ising
  universality class.  In this case the boundary also orders continuously. 
  (This follows from the fact that the $(+)\&(-)$ fixed point is unstable, even
  in zero field, in the ordinary Ising model.)   On the other hand, for smaller
  $\mu$ the bulk transition is first order.  We then expect the boundary
  transition to also be first order since  critical behaviour of the boundary
  at a non-zero $T$ is presumably impossible unless the bulk is also critical. 
  Now consider what happens if $\mu$ is adjusted to its critical value so that
  the bulk transition is in the tri-critical universality class.  In this case
  there are two possibilities for the boundary transition, depending on $J_b$. 
  When $J_b<J_c$, the boundary transition is also second order.  However, when 
  $J_b>J_c$, the boundary undergoes a type of first order transition while the
  bulk undergoes a second order transition.  This follows from observing that,
  infinitesimally below the critical temperature,  the boundary magnetization
  is finite whereas the bulk magnetization is infinitesmal.  On the other hand,
  infinitisimally above $T_c$, the correlation lengths in both bulk and
  boundary are presumably diverging together.  We may understand the
  possibility of the boundary having a first order transition at the bulk
  tri-critical point as being connected with the fact that the bulk system is
  at the end of a first order transition line. When the bulk transition is
  ``almost first order'' it becomes possible for the boundary  transition to be
  truly first order.

  It was recently observed \cite{Recknagel} that the integrable RG flow from 
   $(d)$ to $(+)\&(-)$ 
  has a generalization to all the minimal models
  with diagonal partition functions, which may be thought of as
  increasingly multi-critical Ising models and that, in all cases, the
  $(+)\&(-)$ fixed point is stable except for a dimension 0 operator.  
  This implies that all these models have an ordered phase at zero field, as
  discussed here for the tri-critical case. 
  
  I would like to thank M. Oshikawa and H. Saleur for numerous 
  useful discussions on related subjects and H. Saleur for
  drawing my attention to Ref. \onlinecite{Recknagel}.   
  This research was supported by NSF grant PHY-94-07194 and
  by NSERC of Canada.

\end{document}